\documentclass[fleqn,usenatbib]{mnras}

\usepackage{newtxtext,newtxmath}
\usepackage[T1]{fontenc}

\DeclareRobustCommand{\VAN}[3]{#2}
\let\VANthebibliography\thebibliography
\def\thebibliography{\DeclareRobustCommand{\VAN}[3]{##3}\VANthebibliography}

\usepackage{graphicx}	% Including figure files
\usepackage{amsmath}	% Advanced maths commands
\newcommand{\Porb}{P$_{\rm orb}$}
\newcommand{\msun}{\ensuremath{\, \rm {M}_\odot}}
\newcommand{\rsun}{\ensuremath{\, {R}_\odot}}

\title[Are extreme AGB stars post-common envelope binaries?]{Are extreme AGB stars post-common envelope binaries?}

\author[Dell'Agli et al.]{F. Dell'Agli$^{1}$, E. Marini$^{1,2}$, F. D'Antona$^{1}$, P. Ventura$^{1}$, M. A. T. Groenewegen$^{3}$, 
\newauthor
L. Mattsson$^{4}$, D. Kamath$^{5}$, D. A. Garc\'{\i}a--Hern\'andez$^{6,7}$, M. Tailo$^{8}$
\\
$^1$INAF, Observatory of Rome, Via Frascati 33, 00077 Monte Porzio Catone (RM),           Italy \\
$^2$Dipartimento di Matematica e Fisica, Universit\'a degli Studi Roma Tre, via della Vasca Navale 84, 00100, Roma, Italy \\
$^3$Koninklijke Sterrenwacht van Belgi{\"e}, Ringlaan 3, 1180 Brussels, Belgium \\
$^4$Nordita, KTH Royal Institute of Technology and Stockholm University, Roslagstullsbacken 23, SE-106 91 Stockholm, Sweden\\
$^5$Department of Physics and Astronomy, Macquarie University, Sydney, NSW 2109, Australia\\
$^6$Instituto de Astrof\'{\i}sica de Canarias (IAC), E-38205 La Laguna, Tenerife, Spain\\ 
$^7$Departamento de Astrof\'{\i}sica, Universidad de La Laguna (ULL), E-38206 La Laguna, Tenerife, Spain\\
$^8$Dipartimento di Fisica e Astronomia 'Galileo Galilei', Univ. di Padova, Vicolo dell'Osservatorio 3, I-35122 Padova, Italy\\
}

\pubyear{2015}

\begin{document}
\label{firstpage}
\pagerange{\pageref{firstpage}--\pageref{lastpage}}
\maketitle

% Abstract of the paper
% 250 words for main journal, 200 for letters. ApJ Letters: 250 words
\begin{abstract}
Modelling dust formation in single stars evolving through the carbon-star stage of the asymptotic giant branch (AGB) reproduces well the mid-infrared colours and magnitudes of most of the C-rich sources in the Large Magellanic Cloud (LMC), apart from a small subset of extremely red objects (EROs). The analysis of EROs spectral energy distribution suggests the presence of large quantities of dust, which demand gas densities in the outflow significantly higher than expected from theoretical modelling. We propose that binary interaction mechanisms that involve common envelope (CE) evolution could be a possible explanation for these peculiar stars; the CE phase is favoured by the rapid growth of the stellar radius occurring after C$/$O overcomes unity.
Our modelling of the dust provides results consistent with the observations for mass-loss rates $\dot M \sim 5\times 10^{-4}~\dot M/$yr, a lower limit to the rapid loss of the envelope experienced in the CE phase. We propose that EROs could possibly hide binaries of orbital periods $\sim$days and are likely to be responsible for a large fraction of the dust production rate in galaxies.
\end{abstract}

% Select between one and six entries from the list of approved keywords.
% Don't make up new ones.

%%%%%%%%%%%%%%%%%%%%%%%%%%%%%%%%%%%%%%%%%%%%%%%%%%

%%%%%%%%%%%%%%%%% BODY OF PAPER %%%%%%%%%%%%%%%%%%

\section{Introduction}
Stars of mass in the $\sim 1-8~\rm M_{\odot}$ range evolve through the asymptotic giant branch (AGB) phase. AGB stars experience the third dredge-up (TDU), the inward penetration of the envelope down to regions that previously sited of triple-$\alpha$ nucleosynthesis \citep{iben74}. In stars with mass $< 4 \rm M_{\odot}$ repeated TDU events cause a gradual rise in the surface carbon; if the number of carbon nuclei exceeds that of oxygen, the star becomes a carbon star (C-star). During the AGB evolution the stars are exposed to strong mass loss via cold and dense stellar wind, suitable for the condensation of dust.

A thorough comprehension of the evolution of C-stars is fundamental to understand still poorly known physical mechanisms that take place in stellar interiors, primarily convection and mixing. Furthermore, C-stars are the most efficient sources of carbon gas in the interstellar medium (ISM; \citet{romano20}). 
Furthermore, while in solar-metallicity environments the dust return to the ISM is dominated by O-rich stars \citep{javadi13}, recent studies shown that in lower-metallicity environments the dust production rate (DPR) is dominated by carbon dust \citep[e.g.][]{boyer12, raffa14, flavia16, flavia19}, with a relevant contribution from the most obscured stars.

The existence of stars with extremely red mid-IR colours (EROs) in the LMC
was discovered by \citet{gruendl08}, who used Spitzer IRAC and MIPS photometry and IRS follow-up spectroscopy to identify these objects as extreme C-stars. \citet{sloan16} and \citet{martin18} discussed the peculiarity of these sources and the possible implications related to the large IR emission. \citet{flavia15} and \citet{nanni19} studied the LMC population of evolved stars following an approach based on the results from stellar evolution and dust formation modelling, but could not explain the whole set of photometric and spectroscopic data of the stars in the \citet{gruendl08} sample. 

In this work we reconsider the EROs form the \citet{gruendl08} sample in the attempt of modelling the detailed morphology of the spectral energy distribution (SED), based on IRS data.
We discuss the possibility that their unusual IR excess is connected with the presence of large amounts of dust in their surroundings, due to the common envelope (CE) evolution occurring during the C-star phase of the AGB. We consider binary systems in the orbital period range $\sim$ 2.5-15 yr, where the primary component, an AGB C-star, fills the Roche lobe (RL) first, begins to lose mass and starts a phase of rapid expansion, which further enhances the mass loss rate, a phenomenon that proceeds on a dynamical timescale \citep{paczynski1976}.

This investigation can pose important clues to understand the overall dust production efficiency by evolved stars in galaxies.

\section{Stellar evolution and dust formation modelling}
\label{input}
We calculated evolutionary sequences of $Z=0.008$ stars 
with initial mass in the $1-3.5~\rm M_{\odot}$ range. Stars with mass $M<1~\rm M_{\odot}$ are not considered because they do not become C-stars, as well as $M \geq 4~\rm M_{\odot}$, because they experience hot bottom burning, which destroys the surface carbon and inhibits the C-star stage \citep{ventura14}.

The evolutionary sequences were calculated using the ATON code for stellar evolution \citep{ventura98}. Convection, mass loss, opacities are described as in \citet{ventura14}. We modelled dust formation in the winds of AGB stars following the schematisation proposed by \citet{fg06}. Dust particles are assumed to form and grow in the outflow, which expands isotropically from the photosphere of the star. Dust growth is governed by the rate with which gas molecules collide with pre-existing seed particles, in turn connected to the molecules number density and thermal velocity. Dust formation is assumed to begin at the point in the outflow where the growth rate exceeds the vapourization rate, in turn related to the difference between the
formation enthalpies of the solid compound and the individual gaseous reactants.
All the relevant equations can be found in \citet{ventura12}. 

\begin{figure*}
\begin{minipage}{0.33\textwidth}
\resizebox{1.\hsize}{!}{\includegraphics{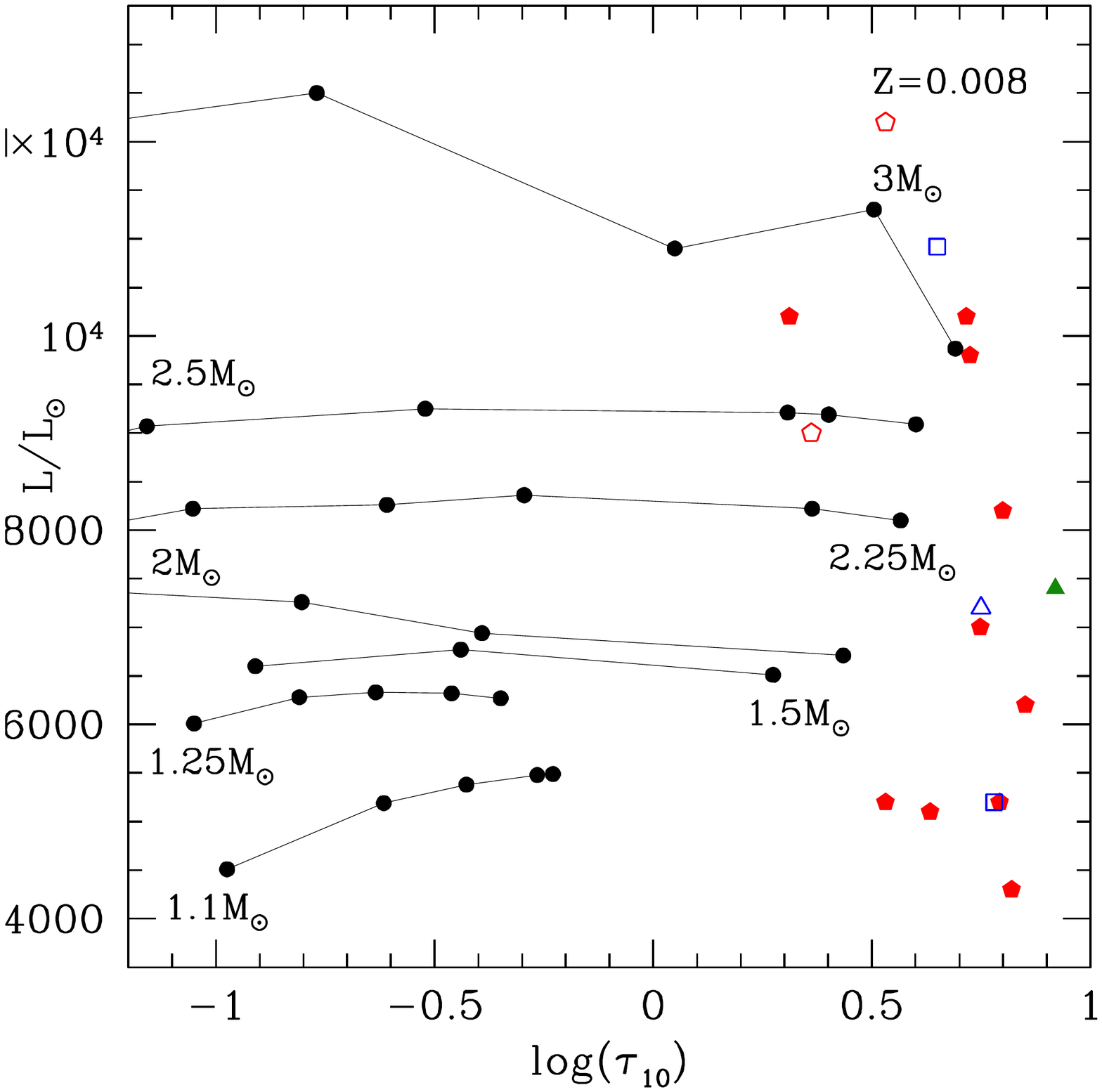}}
\end{minipage}
\begin{minipage}{0.33\textwidth}
\resizebox{1.\hsize}{!}{\includegraphics{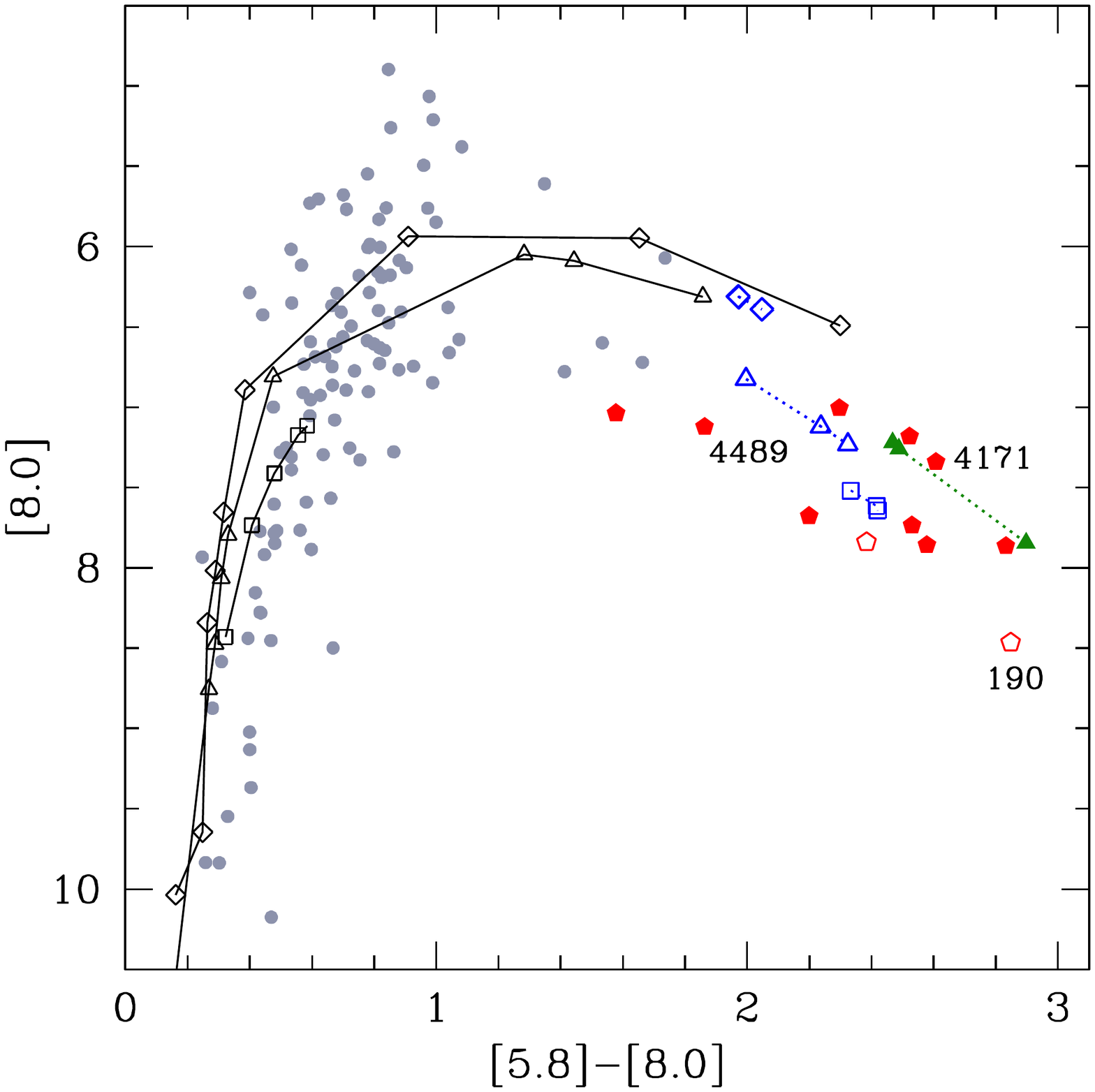}}
\end{minipage}
\begin{minipage}{0.33\textwidth}
\resizebox{1.\hsize}{!}{\includegraphics{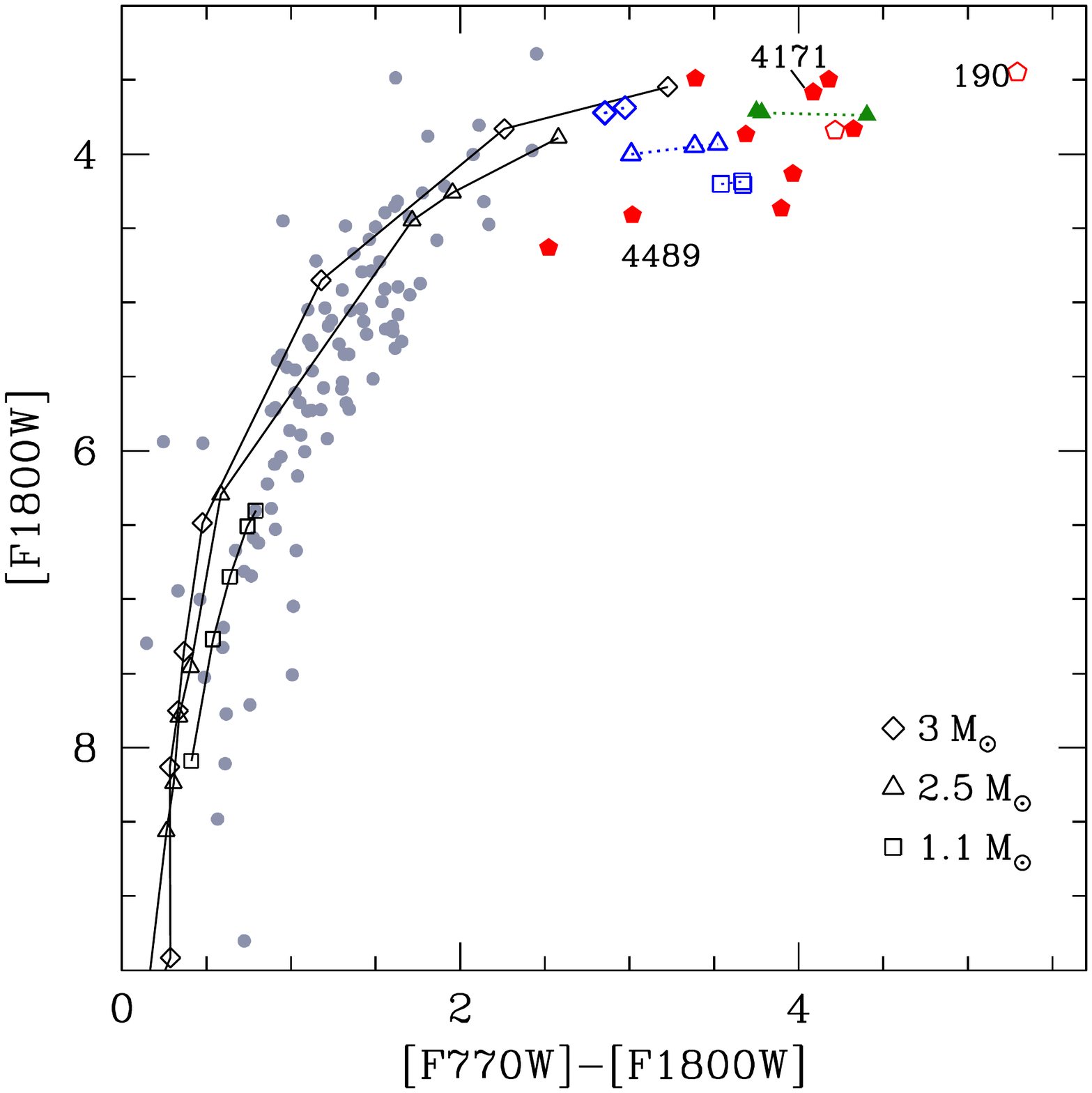}}
\end{minipage}
\vskip-40pt
\caption{Left: The evolution of stars of different mass in the $\tau_{10}- \rm L/L_{\odot}$ plane. Red pentagons indicate the optical depths and magnitudes of the EROs sources, obtained by SED fitting. Open blue points refer to the evolution of 1.1$~\rm M_{\odot}$ (square), 2.5$~\rm M_{\odot}$ (triangle) and 3.0$~\rm M_{\odot}$ (diamond) when assuming an enhanced mass-loss rate immediately after the beginning of the C-star phase or in a more advanced stage (full green triangles). Central and right: The evolutionary tracks (black solid lines), LMC C-stars from the SAGE-Spec database (grey points) and EROs sources (red pentagons) in the colour-magnitude ([5.8]--[8.0], [8.0]) and ([F770W]--[F1800W], [F1800W]) planes, respectively. The evolutionary points found when assuming an enhanced mass-loss rate (blue points with dashed lines), simulating the CE phase, are reported in both panels. Open pentagons refer to SSID 190 and SSID 125.}
\label{fccd}
\end{figure*}

The dust formation modelling is applied to some points during each inter-pulse and thermal pulse (TP) phase, chosen to properly follow the variation in the evolutionary properties of each star. This allows to determine the evolution of the dust composition and of $\tau_{10}$, the optical depth at $\lambda = 10~\mu$m. The outcomes of the star+dust modelling were used as input for the radiation transport code DUSTY \citep{nenkova99}, to build a sequence of synthetic SEDs, which allow to simulate the evolution of the (circum)stellar spectrum. The convolution with the transmission curves of various filters allows to compute evolutionary tracks in the observational planes based on different photometric systems (black tracks in Fig.~\ref{fccd}).

\section{The evolutionary pattern of carbon stars}
\label{model}
The left panel of Fig.~\ref{fccd} shows the evolution of AGB star models in the $\tau_{10}-L$ plane.
The increase in $\tau_{10}$ is due to the gradual rise in the surface
carbon and the consequent expansion of the external regions \citep{vm09}. These factors lead to a more efficient dust formation, owing to the higher availability of carbon molecules and to the cooler temperatures in the envelope, which favour condensation with respect to vaporisation \citep{fg06}. 

\begin{figure*}
\begin{minipage}{0.33\textwidth}
\resizebox{1.\hsize}{!}{\includegraphics{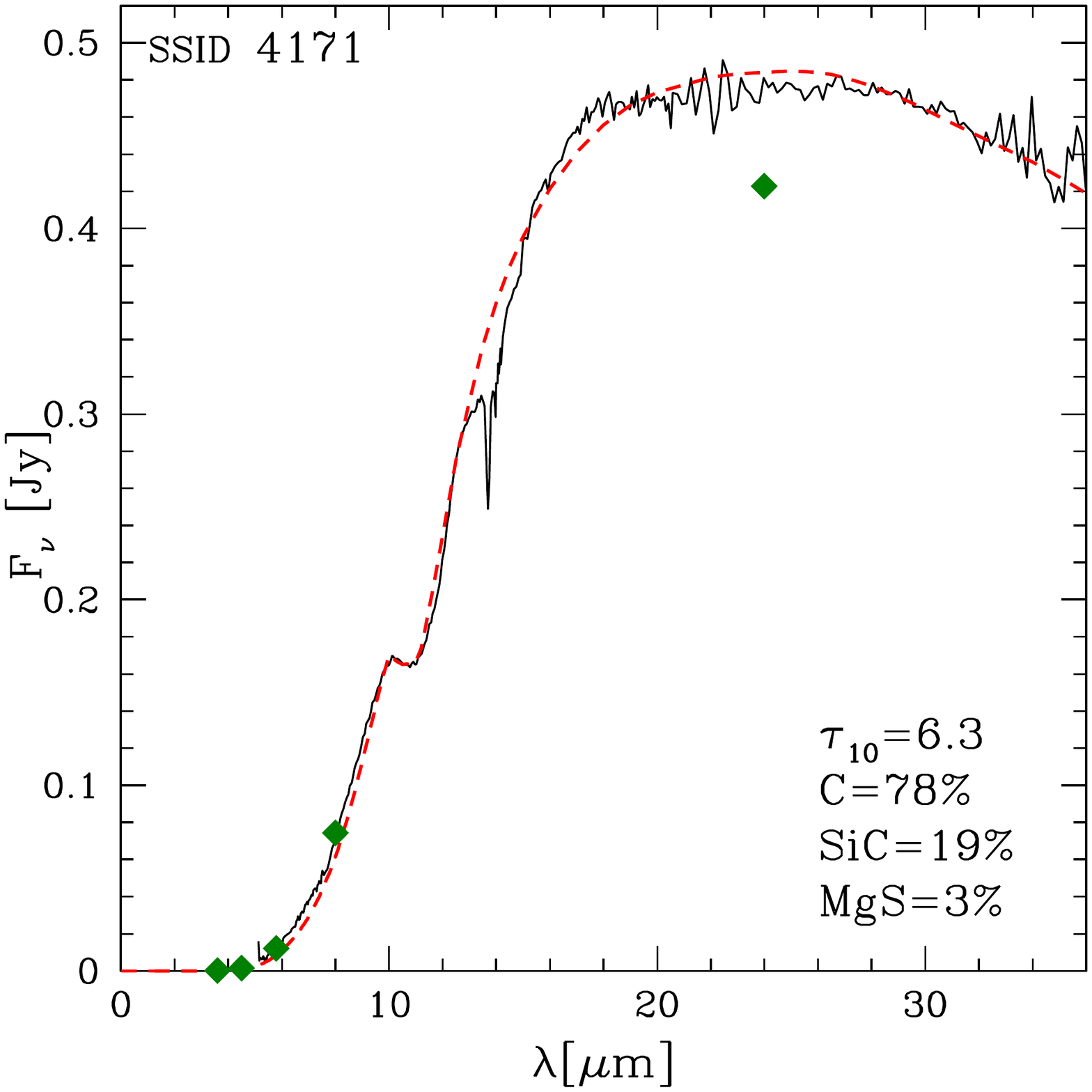}}
\end{minipage}
\begin{minipage}{0.33\textwidth}
\resizebox{1.\hsize}{!}{\includegraphics{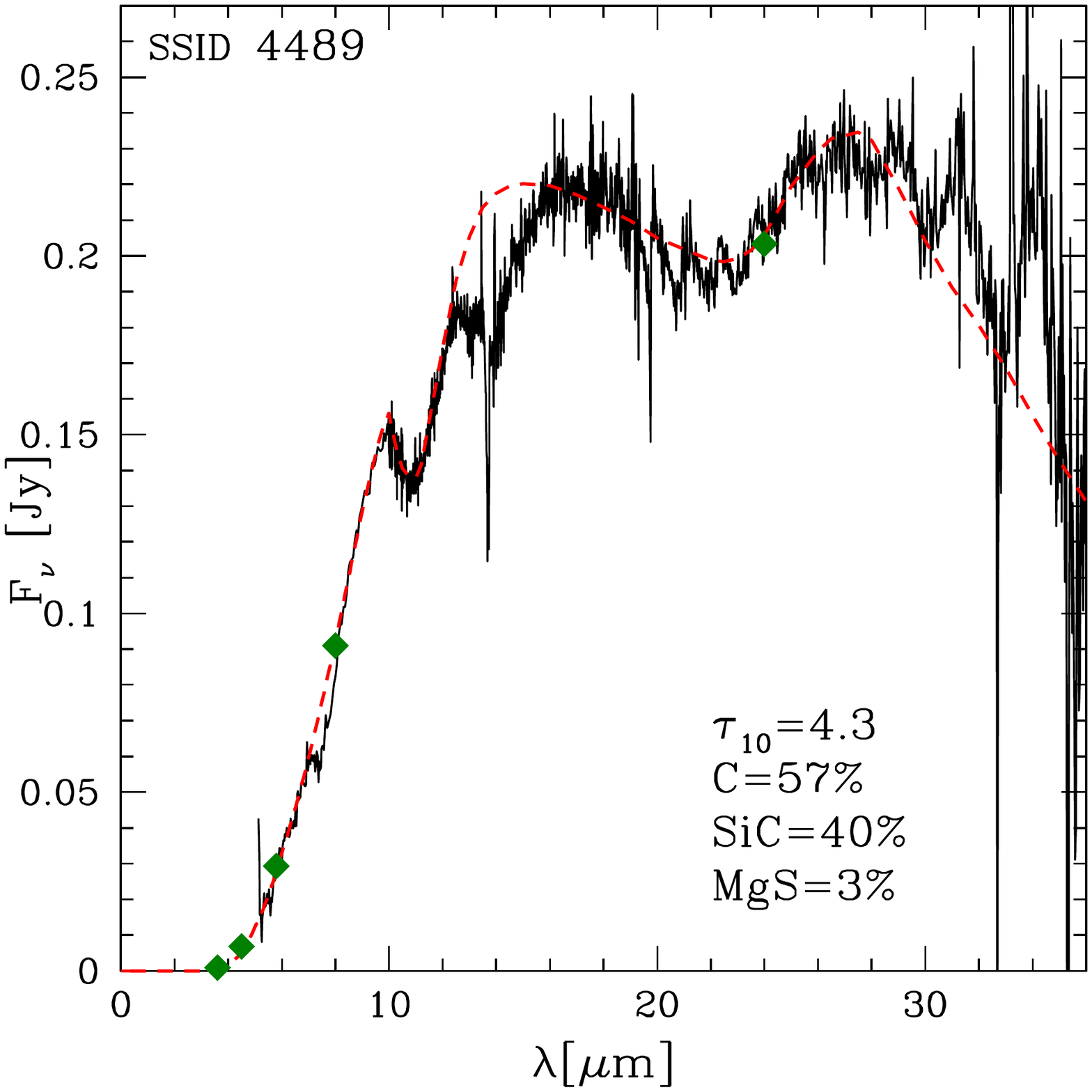}}
\end{minipage}
\begin{minipage}{0.33\textwidth}
\resizebox{1.\hsize}{!}{\includegraphics{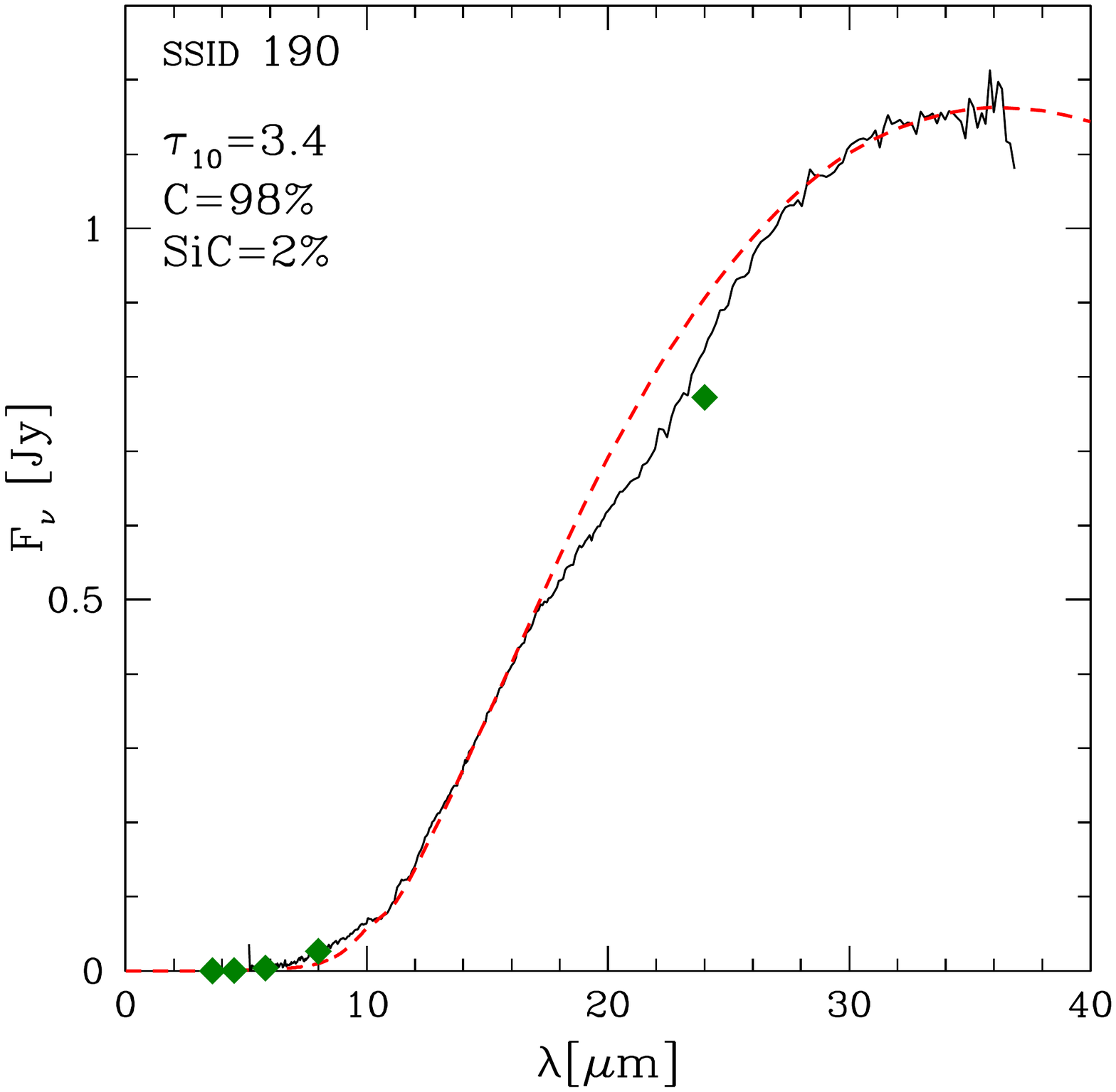}}
\end{minipage}
\vskip-40pt
\caption{IRS SED of SSID 4171 (left panel), SSID 4489 (centre) and SSID 190 (right) indicated with black lines and the best fit (red lines), obtained with the optical depth and the dust mineralogy reported in the figures. Photometry from IRAC and MIPS is indicated with green diamonds.}
\label{fsed}
\end{figure*}

In the central panel of Fig.~\ref{fccd} we show the evolutionary tracks in the ([5.8]--[8.0], [8.0]) colour-magnitude diagram constructed on the basis of the IRAC filters of the \textit{Spitzer Space Telescope}. The right panel of Fig.~\ref{fccd} shows the evolutionary tracks in the ([F770W]--[F1800W], [F1800W]) diagram, where [F770W] and [F1800W] are the magnitudes obtained by convolving the synthetic SED with the transmission curves of two mid-IR filters of the MIRI camera, mounted onboard the upcoming \textit{James Webb Space Telescope (JWST)}. 

\subsection{The extreme carbon star stage}
Grey points and red pentagons in Fig.~\ref{fccd} refer to the
LMC C-stars observed by \textit{Spitzer}, in the area of the sky covered by the SAGE survey \citep{meixner06}. \citet{jones17} calculated the expected [F770W] and [F1800W] of these objects, via convolution of the IRS spectra with the MIRI transmission curves. 

Here we focus on the stars characterised by large IR excess, indicated with red symbols in Fig.~\ref{fccd}, that
involve the extreme stars from \citet{gruendl08} and SSID 9, which shares similar characteristics. The appropriate combination of $\tau_{10}$ and dust composition allows to fit the observations, by reproducing all the major features in the spectra. The position of the red points in the left panel of Fig.\,\ref{fccd} was determined by adopting this method. Two examples of the SED fitting results are shown in the left and central panel of Fig.\,\ref{fsed} for the sources SSID 4171 and 4489. The main properties derived for the EROs are summarised in Table \ref{tabero}.

\begin{table*}
\caption{Summary of the interpretation of the stars discussed in this work: Spitzer and IRAC/ERO name, coordinates, luminosity, optical depth, percentages of the different dust species (in order: solid carbon, silicon carbide, magnesium sulphide, graphite), initial mass and age. }    
\label{tabero}      
\centering          
\begin{tabular}{c c c c c c c c c c c c}     
\hline       
SSID & IRAS/ERO name & RA (deg) & DEC (deg) & L$/$L$_{\odot}$ & $\tau_{10}$ & $\%($C) & $\%($SiC) & $\%($MgS) & $\%($graph) & M$_{\rm init}/$M$_{\odot}$ & Age (Gyr) \\ 
\hline                    
  4185 & IRAS 05042-6827 & 76.0233 & $-$68.3945 & 5200  & 6.2 & 67 & 25 & 2 & 6  & $1.1-1.2$ & $5.0-7.0$     \\ 
  4299 & IRAS 05187-7033 & 79.5488 & $-$70.5075 & 9800  & 5.3 & 79 &  0 & 0 & 21 & $2.5-3.0$   & $0.4-0.6$ \\
  4308 & IRAS 05191-6936 & 79.7016 & $-$69.5596 & 7000  & 5.6 & 64 & 22 & 2 & 12 & $2.0-2.5$   & $0.6-1.0$   \\
  4415 & IRAS 05260-7010 & 81.4193 & $-$70.1409 & 4700  & 6.6 & 68 & 18 & 2 & 12 & $1.1-1.2$ & $5.0-7.0$     \\
  4171 & ERO 0502315     & 75.6312 & $-$68.0934 & 8200  & 6.3 & 68 & 19 & 3 & 10 & $2.0-2.5$   & $0.6-1.0$   \\
  4489 & IRAS 05305-7251 & 82.4079 & $-$72.8314 & 5100  & 4.3 & 53 & 40 & 3 & 4  & $1.1-1.2$ & $5.0-7.0$     \\
  4781 & IRAS 05509-6956 & 87.6091 & $-$69.9342 & 10200 & 5.2 & 68 & 27 & 3 & 2  & $2.5-3.0$   & $0.4-0.6$ \\
  9    & IRAS 04518-6852 & 72.9192 & $-$68.7930  & 5200  & 3.4 & 75 & 14 & 5 & 6  & $1.1-1.2$ & $5.0-7.0$     \\ 
  65   & IRAS 05133-6937 & 78.2576 & $-$69.5642 & 6200  & 7.1 & 76 & 20 & 2 & 2  & $1.5-2.0$   & $1.0-2.0$     \\
  125  & IRAS 05315-7145 & 82.6853 & $-$71.7167 & 9000  & 2.3 & 62 & 14 & 2 & 24 & $2.0-2.5$   & $0.6-1.0$   \\
  190  & IRAS 05495-7034 & 87.2504 & $-$70.5562 & 12200 & 3.4 & 83 &  2 & 2 & 15 & $3.0-3.5$   & $0.3-0.4$ \\
\hline

\end{tabular}
\end{table*}

The optical depths of these stars are significantly higher than the theoretical expectations. Some consistency is found only for the brightest extreme stars, whose luminosity and $\tau_{10}$
are similar to those of the last point of the track of the $3~\rm M_{\odot}$ star.
In the colour-magnitude plane, the evolutionary tracks reproduce the position of most evolved LMC stars in the sample, with the exception of the EROs.

The method used to determine the optical depths of the model described in Sect.\,\ref{input} has some limitations. The stationary wind neglects the effects of pulsation \citep{hoefner18}, which might potentially eject dense gas clouds into cool regions of the circumstellar envelope, where dust formation would be favoured. Furthermore, no gas-dust drift is considered, which might lead to an underestimation of the dust formed \citep{sandin20}. 

Despite these uncertainties, we believe that there is not much room to significantly widen the extension of the evolutionary tracks in Fig.~\ref{fccd}. The key factor for dust formation in the outflow during a given evolutionary stage is the mass-loss rate $\dot{\rm M}$, which determines the density of the wind and thus of the molecular species available to condensation. The terminal points of the sequences in Fig.~\ref{fccd} are characterised by $\dot{\rm M} \sim 1-2\times 10^{-4}~\rm M_{\odot}/$yr, calculated based on \citet{wachter02}, which might overestimate the $\dot{\rm M}$ of C-stars \citep{bladh19}. Higher $\dot{\rm M}$ are thus difficult to justify in the present context. The discrepancy between modelling and observations is particularly relevant for the three faintest stars, whose luminosities indicate $\sim 1-1.5~\rm M_{\odot}$ progenitors. These stars reach the C-star stage during the final AGB phases, with a surface C$/$O below 1.5: the corresponding carbon excess with respect to oxygen is likely not sufficient to drive a dust-driven wind \citep{bladh19}.

\subsection{The effects of enhanced mass-loss}
\label{massloss}
We tested the effects on dust production if we increase the mass loss rate at an earlier evolutionary stage, namely soon after C/O$>$1. We resumed the computations of the 1.1, 2.5 and $3~\rm M_{\odot}$ models, from the stages indicated by the blue arrows in the left panel of Fig.~\ref{fperiod}, after the jump in the radius following a TDU episode. We kept the mass loss rate at $\dot{\rm M} = 5\times 10^{-4}~\rm M_{\odot}/$yr, a first order approximation to the situation encountered in the CE of a binary system during the evolution (see next Section), where an envelope of $\sim 1$\msun\ can be lost even in 10\,yr \citep{chamandy2020} and matter will transfer out of the Lagrangian points until it is dispersed. For the $2.5~\rm M_{\odot}$ model star we considered an additional case, where the enhanced mass loss is assumed from a later stage (green arrow in Fig.~\ref{fperiod}). We model dust
formation as in the spherical, single star evolution (SSE), although the time-scale of the events is
so short that the details of the nucleation processes which lead to the formation of the
seed particles on which dust grains grow might be different. Besides, neglecting the non-spherical structure of the outflow might lead to an overestimation of the $\dot{\rm M}$ required to account for the observed IR emission \citep{sloan16,wiegert20}. These points deserve further 
investigation in the future. Despite this, CE evolution appears as a good candidate to justify the dust emission of these EROs, because the higher mass-loss favoured by the CE interaction increases the density of gaseous molecules available to form dust, consistently with the study by \citet{glanz18}, who showed that efficient dust production takes place follwing CE interaction.

The evolution of the  
tracks in Fig.\,\ref{fccd} during the enhanced mass loss phase (blue and green points) reproduces the observed locations of EROs. The reddest objects are better reproduced
by the models where the enhanced mass loss is started later (green triangles), because the higher surface carbon favours a larger dust production.
In the $3~\rm M_{\odot}$ case use of the higher $\dot M$ does not change $\tau_{10}$ in a meaningful way. This is because the surface carbon mass fraction at the point when we assumed the large $\dot M$ is 3 times smaller than during the final phases of the "standard" case, corresponding to the reddest point of the evolutionary track; this effect
counterbalances the higher $\dot M$. This is also the reason why the dust yields of $2.5-3~\rm M_{\odot}$ stars are unchanged with respect to the SSE (see Fig. 10 in \citet{ventura14}). Conversely, in $1.1~\rm M_{\odot}$ stars, in which the CE phase begins towards the end of the AGB, the C dust produced is two times higher.

 The spectra of SSID 125 and 190 (see Fig.~\ref{fsed} for the latter source) show up dissimilarities with the other sources discussed here. From the interpretation of their SED we find dust temperatures $\sim 300$ K, much cooler than the other stars, and optical depths $\tau_{10} \sim 2-3$, the lowest reported in Table \ref{tabero}. These results might indicate
that the primary has started the post-AGB evolution, with the dusty region moving away from the central star \citep{martin18}.

\section{Common envelope evolution: how to reach high mass loss at an earlier phase}
In the previous Section we describe how the large IR emissions can be obtained via the occurrence of very large mass loss rates  earlier during the AGB phase, even as soon as C/O$>$1. In this Section, we explore a plausible situation which may produce this effect.

In fact, if the AGB star is in a binary system, its evolution may be affected by the presence of the companion. The basic subdivision of binary interaction made by \cite{paczynski1971} classifies as "Case C" the evolution of a binary when the primary component (the most massive star) fills its RL and begins mass exchange during the AGB phase. As a reaction to mass loss, the stellar radius of a donor having a deep convective envelope increases, while the binary separation decreases when mass is transferred to the less massive companion. Thus, mass is lost from the AGB star at very high rates, while the cores of the two components spiral towards each other in a CE phase \citep{paczynski1976}. Note that pre-CE binary interaction has generally
negligible consequences for the CE event. The binary interaction occurs when the systems have orbital period \Porb\ of years, but the orbital angular momentum transfer from the inner cores to the outer layers is so large that the final period may be in the range of days or even hours \citep[e.g.][]{willemskolb2004}, and the shortest \Porb\ post-CE systems will be the progenitors of cataclysmic binaries (CBs), in which a low mass quasi-main sequence star (the previous secondary star) is now the component filling the RL and transferring mass to the (former primary) white dwarf (WD). The existence of binary nuclei of planetary nebulae \cite[see e.g.][]{jonesboffin2017} has confirmed the scheme, and the properties of the remnant post-CE binary and CB population have been explored in several population synthesis models \citep[e.g.][]{willemskolb2004}\footnote{Not all post--CE binaries evolve into the interacting stage, and research is now ongoing \citep[e.g.][]{wijnen2015} to understand why the average mass of the WD components of CBs are much larger ($\sim 0.8$\msun) than the average mass of carbon-oxygen white dwarfs remnants of post--CE binaries ($\sim 0.6$\msun) \citep{zorotovic2011}, which is also the typical WD mass remnant of our proposed evolution.}. 

Full consideration of the effect of the wind mass loss from the AGB star and of the structure details of the AGB envelopes and the effect of TPs may be important to understand which systems evolve into stable case C evolution \citep{pastetter1989} and to outline the precise evolutionary boundaries for which the present proposal is feasible. However, for this study we are interested mainly in those systems which: 1) suffer unavoidable CE evolution; 2) when they are already C-stars; 3) where the CE allows to meet the required dust density production which is not allowed during the SSE. 

\subsection{The stellar radius evolution}
\begin{figure*}
%\begin{minipage}{0.48\textwidth}
\resizebox{0.43\hsize}{!}{\includegraphics{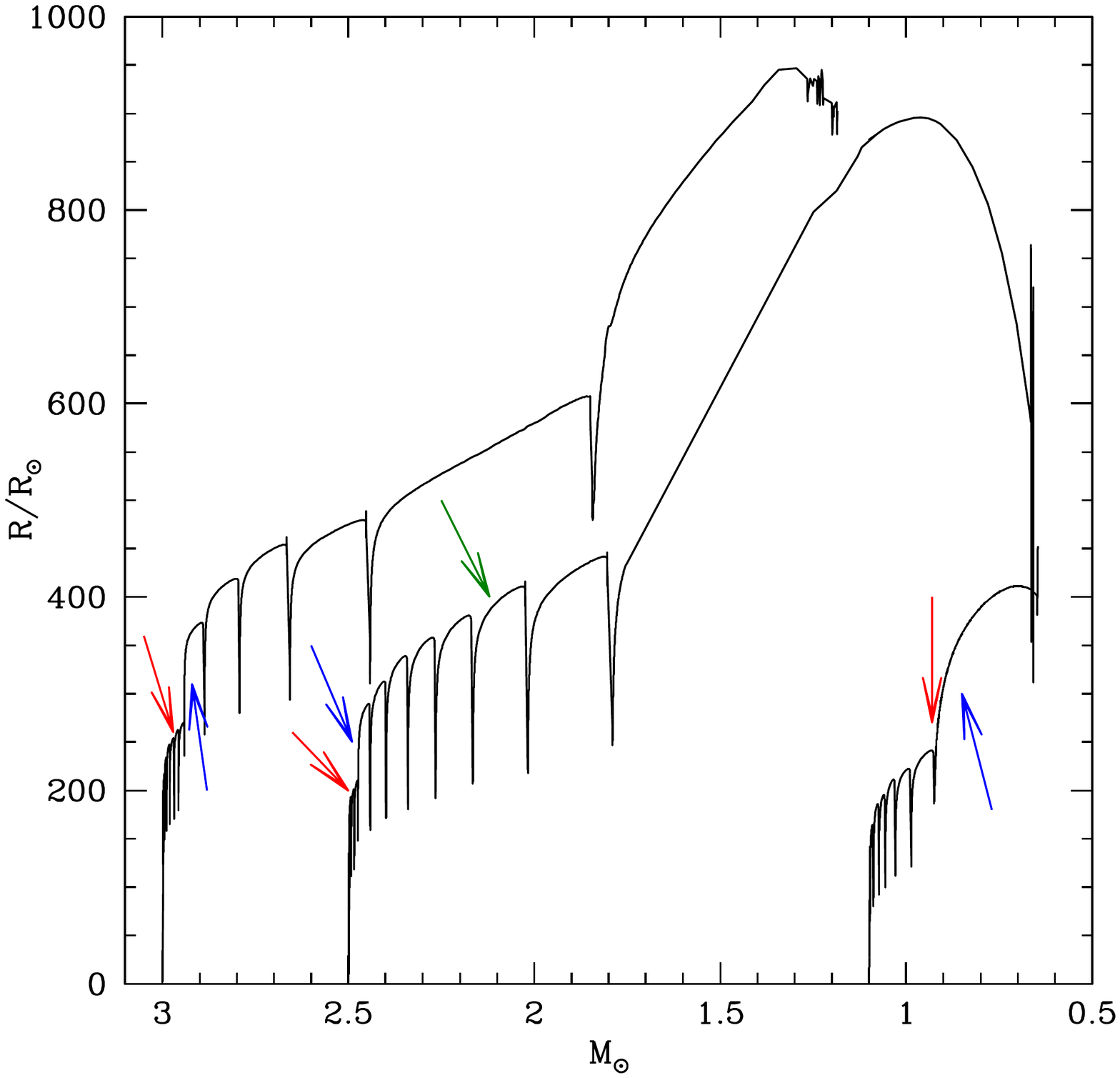}}
%\end{minipage}
%\begin{minipage}{0.48\textwidth}
\resizebox{0.43\hsize}{!}{\includegraphics{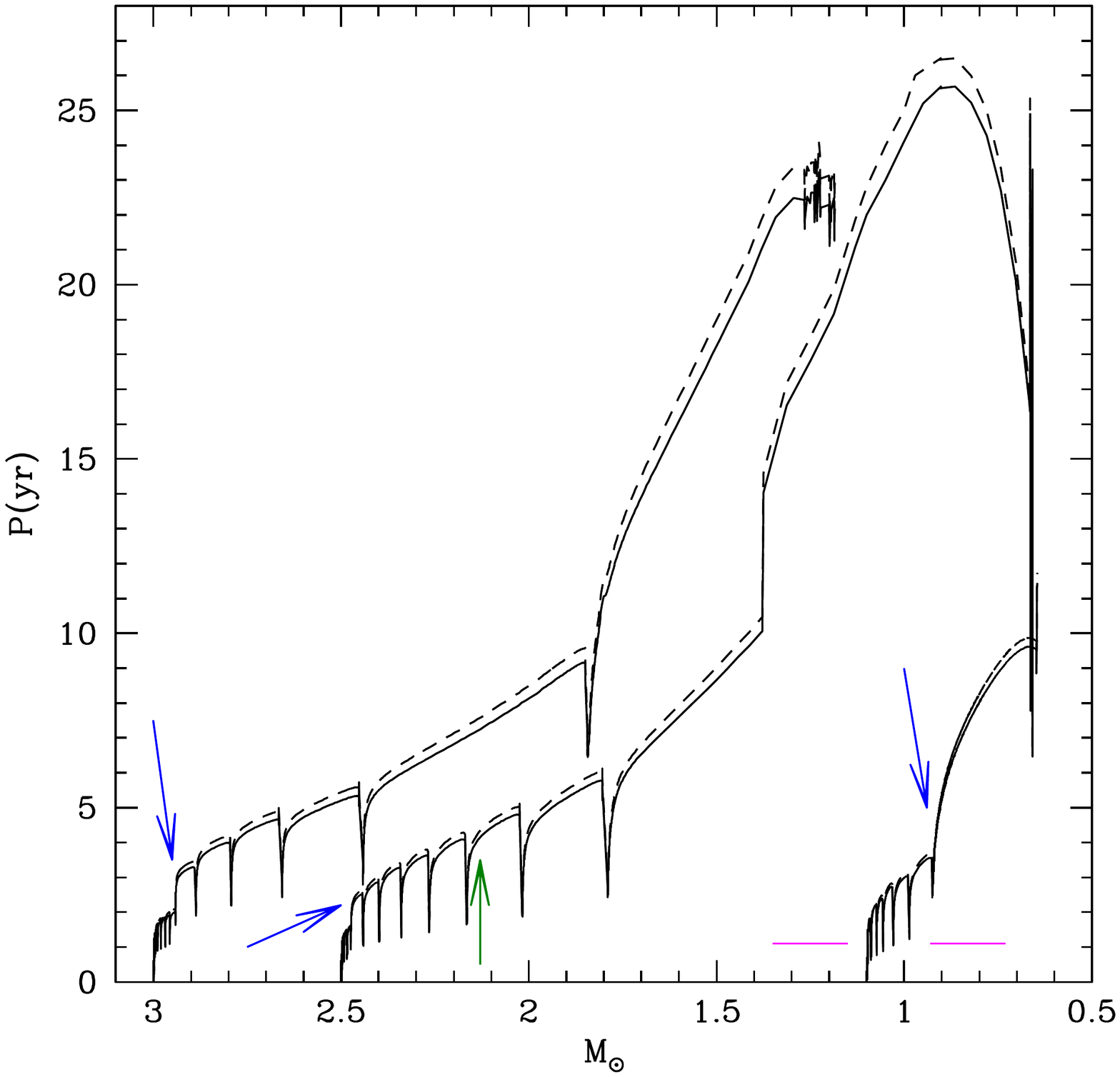}}
%\end{minipage}
\vskip-60pt
\caption{Left: variation of the radius of initial mass models
$1.1~\rm M_{\odot}$, $2.5~\rm M_{\odot}$ and $3~\rm M_{\odot}$, as a function of the
current mass of the star. Arrows mark the beginning of the C-star phase (red) and when the star rapidly expands (blue and green), due to the surface carbon enrichment. Right: orbital period evolution, intended as the initial period of a binary in which the AGB fills the RL at each point of mass along its SSE. Solid and
dashed lines refer to the cases in which the companion mass is $0.6~\rm M_{\odot}$ and $0.8~\rm M_{\odot}$, respectively. If contact occurs at the indicated points (blue or green arrows), a CE evolution and rapid shrinking of the period is expected. Magenta line marks the period of the system if the contact
would occur when the $1.1~\rm M_{\odot}$ star is at the tip of the RGB.
}
\label{fperiod}
\end{figure*}

The main parameter determining whether the AGB star will evolve into RL contact is the evolution of the stellar radius. In our case, the current binary separation (or \Porb) must be such that mass exchange did not begin during the red giant branch (RGB) phase. We examine three representative masses: 1.1, 2.5 and 3.0\msun.
The left panel of Fig.~\ref{fperiod} shows the radius evolution as a function of the current mass for the three cases. Note that the SSE includes mass loss by stellar wind, so the mass is a proxy of the advancement of the time. 

While the 2.5 and 3\msun\ cases have relatively small radii at the onset of central helium burning (29 and 34\rsun), the 1.1\msun\ undergoes helium flash at a radius $R=124$\rsun; this value is smaller than during the first TPs and $\sim 2.5$ times lower than during the C-star phase. We consider binary systems exceeding the \Porb\ $\sim  1$ yr, which would have produced mass exchange during the RGB. 

As shown in Fig.~\ref{fperiod}, the radius during the SSE begins increasing with a steep derivative for C$/$O$\sim1.5$, owing to the increase in the surface molecular opacities (see Sect. \ref{model}). Thus, the case in which the AGB star is the primary in a detached binary, filling the RL as C--star, occurs for a wide range of \Porb. In the right panel of Fig.\,\ref{fperiod} is shown \Porb\ for a (non interacting) binary in which the primary has the AGB mass, with a radius equal to the RL radius \citep[computed following eq. 2 and 3 in ][ see left panel of Fig.\ref{fperiod}]{eggleton1983}, and the companion has a mass of 0.6\msun  or 0.8\msun. \Porb\ in Fig.\,\ref{fperiod} is thus the ``initial" possible period of a binary at the moment it can interact with the companion. It is self-evident that, if the primary has not filled the RL in previous evolutionary phases, there is a huge range of initial periods ($\sim$ $2.5-15$ yrs, the precise range depending on the evolving mass) during which this may happen, giving origin to the high mass loss which favours the production of dust. 

\section{Discussion and conclusions}
We propose that the EROs can be interpreted as the result of the evolution of binaries of periods $\sim$2.5--15\,yr (see Fig.~\ref{fperiod}), in which the primary is an AGB star of mass 1.1--3.0\msun\ evolving through the C-star phase, and the companion is a star of mass low enough that the mass transfer is unstable.  $M>1.2~\rm M_{\odot}$ companion stars cannot be considered, as they expand as a reaction to mass accretion, and the system would follow a different route with respect to the one proposed here. The binary system evolves through a CE phase and ends up as a pre--cataclysmic binary of periods of the order of days. 
We simulate the evolution of the AGB by assuming a high constant mass loss rate, and computing the dust evolution. 
This work provides proof of concept for the hypothesis that the parameters of the CE evolution are not particularly tight, and that the resulting dust is indeed of density high enough to produce the colors of the EROs. Computation of populations synthesis may be useful to constrain the number of such systems expected in the Galaxy; these stars may be more efficient dust producers than the much more numerous AGBs evolving as single stars \citep[e.g.][]{boyer12}. The fact that the EROs observed so far are all C-stars can be explained by the constrains given above on the mass of the companion: 
systems with a $4-6~\rm M_{\odot}$ primary would experience a CE evolution only if the 
mass ratio is above 4-5, which renders these events rather rare.
We conclude this work with a plea to look for signs of carbon dust, having survived the CE ejection, in post-CE systems where the AGB descendant has 0.55$\leq \rm M/M_{\odot}\leq$0.7, typical core mass of C-stars.

%% Example figure
%\begin{figure}
%	% To include a figure from a file named example.*
%	% Allowable file formats are eps or ps if compiling using latex
%	% or pdf, png, jpg if compiling using pdflatex
%	\includegraphics[width=\columnwidth]{example}
%   \caption{This is an example figure. Captions appear below each figure.
%	Give enough detail for the reader to understand what they're looking at,
%	but leave detailed discussion to the main body of the text.}
%   \label{fig:example_figure}
%\end{figure}

%% Example table
%\begin{table}
%	\centering
%	\caption{This is an example table. Captions appear above each table.
%	Remember to define the quantities, symbols and units used.}
%	\label{tab:example_table}
%	\begin{tabular}{lccr} % four columns, alignment for each
%		\hline
%		A & B & C & D\\
%		\hline
%		1 & 2 & 3 & 4\\
%		2 & 4 & 6 & 8\\
%		3 & 5 & 7 & 9\\
%		\hline
%	\end{tabular}
%\end{table}

\section*{Acknowledgements}
We would like to thank the referee Dr. J. Th. van Loon for his useful suggestions. 
DAGH acknowledges support from the State Research Agency (AEI) of the 
Spanish Ministry of Science, Innovation and Universities (MCIU) and 
the European Regional Development Fund (FEDER) under grant 
AYA2017-88254-P. DK acknowledges the support of the Australian Research Council (ARC) Decra grant (95213534).

\section*{Data Availability}
The data underlying this article will be shared on reasonable request
to the corresponding author.
 
%The inclusion of a Data Availability Statement is a requirement for articles %published in MNRAS. Data Availability Statements provide a standardised format %for readers to understand the availability of data underlying the research %results described in the article. The statement may refer to original data %generated in the course of the study or to third-party data analysed in the %article. The statement should describe and provide means of access, where %possible, by linking to the data or providing the required accession numbers for %the relevant databases or DOIs.

%%%%%%%%%%%%%%%%%%%% REFERENCES %%%%%%%%%%%%%%%%%%

% The best way to enter references is to use BibTeX:

\bibliographystyle{mnras}
\bibliography{example} % if your bibtex file is called example.bib

% Alternatively you could enter them by hand, like this:
% This method is tedious and prone to error if you have lots of references
%\begin{thebibliography}{99}
%\bibitem[\protect\citeauthoryear{Author}{2012}]{Author2012}
%Author A.~N., 2013, Journal of Improbable Astronomy, 1, 1
%\bibitem[\protect\citeauthoryear{Others}{2013}]{Others2013}
%Others S., 2012, Journal of Interesting Stuff, 17, 198
%\end{thebibliography}

%%%%%%%%%%%%%%%%%%%%%%%%%%%%%%%%%%%%%%%%%%%%%%%%%%

%%%%%%%%%%%%%%%%% APPENDICES %%%%%%%%%%%%%%%%%%%%%

%\appendix

%\section{Some extra material}

%If you want to present additional material which would interrupt the flow of the %main paper, it can be placed in an Appendix which appears after the list of %references.

%%%%%%%%%%%%%%%%%%%%%%%%%%%%%%%%%%%%%%%%%%%%%%%%%%

% Don't change these lines
\bsp	% typesetting comment
\label{lastpage}
\end{document}